# Proton Transfer in Phase IV of Solid Hydrogen and Deuterium

Hanyu Liu and Yanming Ma[*]

*State Key Laboratory of Superhard Materials, Jilin University, Changchun 130012, China*

The recent discovery of phase IV of solid hydrogen and deuterium consisting of two alternate layers of graphene-like three-molecule rings and unbound $H_2$ molecules have generated great interests. However, vibrational nature of phase IV remains poorly understood. Here, we report a peculiar proton transfer and a simultaneous rotation of three-molecule rings in graphene-like layers predicted by *ab initio* variable-cell molecular dynamics simulations for phase IV of solid hydrogen and deuterium at pressure ranges of 250 – 350 GPa and temperature range of 300 - 500 K. This proton transfer is intimately related to the particular elongation of molecules in graphene-like layers, and it becomes more pronounced with increasing pressure at the course of larger elongation of molecules. As the consequence of proton transfer, hydrogen molecules in graphene-like layers are short-lived and hydrogen vibration is strongly anharmonic. Our findings provide direct explanations on the observed abrupt increase of Raman width at the formation of phase IV and its large increase with pressure.

A new insulating phase IV of solid hydrogen was observed above 220 GPa transformed from phase I (disorder *hcp*) at room temperature by two recent Raman experiments [1, 2]. This phase IV [2] was interpreted as having *Pbcn* symmetry and 24 molecules per unit cell, which was earlier proposed by first-principles structure predictions as a possible candidate for phase III at zero kelvin [3]. This unusual structure includes two alternate layers of graphene-like three-molecule rings with elongated $H_2$ molecules and unbound $H_2$ molecules (see Fig. 1) [2]. However, this *Pbcn* structure was calculated to have severely imaginary (unstable) phonons and a slightly improved *Pc* structure with more distorted graphene-like sheets was proposed to eliminate the imaginary phonons [4]. Our first-principles metadynamics studies [5] proposed another approximate *Cc* structure containing 96 molecules per cell and suggested phase IV is in fact a partially disordered structure: disordering in weakly-coupled or unbounded $H_2$ molecules layer (layer I, Fig. 1b), but ordering in alternate strongly-coupled graphene-like layers (layer II, Fig.1c). The disordering of phase IV is apparently related to the high-temperature nature in its existence. To note, low-temperature phase III transforms to phase IV at 270 K and 255 GPa [2]. Entropy-driven formation of phase IV is clear [4, 5].

Zero-kelvin density functional theory studies proposed that graphene-like layers with elongated $H_2$ molecules can dissociate into perfect atomic graphene layers above 350 GPa [4]. Once it is stabilized, the atomic layer may possess quantum liquid behaviors due to the large zero-point vibrations [2, 4]. As a result, unusual physical phenomena can be expected [2]. To date, previous *ab initio* path-integral molecular dynamics (MD) simulations [6] by using a supercell of 64 molecules were able to propose that hydrogen in a hypothetic string-like molecular structure can diffuse or experience quantum tunneling at 350 GPa, though their simulations performed were not using variable-cell MD with fully converged k-mesh (only gamma point sampling was employed). In recent experiments [1, 2], it was observed that full width at half maximum (FWHM) of the $v_1$ Raman peaks has an abnormally abrupt increase at the formation of phase IV, and it further increases largely with increasing pressure. These experimental studies [1, 2] implied phase IV has rather unusual vibrational properties, which unfortunately were poorly understood. Further vibrational studies of phase IV are thus greatly desirable.

Here, we explored the vibrational properties of phase IV of solid hydrogen and deuterium at temperatures of 300 – 500 K and in the pressure range of 250-350 GPa by *ab initio* variable-cell MD simulations. We have unambiguously established that hydrogen protons in the graphene-like layer of phase IV can readily transfer to their neighboring molecular sites by accompanying a simultaneous rotation of three-molecule rings. Proton transfer becomes more pronounced at elevated pressures. Our simulations provided direct evidences on the short lifetime of hydrogen molecules and strong an-harmonic vibration of hydrogens in the graphene-like layer of phase IV. Our findings fundamentally explain the abrupt increase of $v_1$ Raman FWHM and its increase with pressure in phase IV.

The *ab initio* MD simulations for solid hydrogen and deuterium were performed within the *NPT* (*N*-number of particles, *P*-pressure, *T*-temperature) ensemble [7], as recently implemented in Vienna *ab initio* Simulation Package (VASP) code [8-10]. The all-electron projector-augmented wave (PAW) [11, 12] method was adopted. Exchange and correlation effects were treated in generalized-gradient approximation (GGA) [13]. We have adopted a simulation system containing 48 molecules for phase IV and the simulation times (>7 ps) are typically long. A plane-wave cutoff of 500 eV and the Brillouin zone sampling with a 4×4×4 *k*-mesh were employed [14]. The time step in the MD simulations was 0.5 fs, and the self-consistency on the total energy was chosen to be $2\times10^{-5}$ eV. Our MD simulations have been validated by a direct modeling of the observed temperature-induced phase III→IV transformation at 250 GPa. Our MD runs started with the $C2/c$ structure [3] for phase III by employing a simulation cell of 48 molecules. The $C2/c$ structure remained very stable after 5 ps simulations at temperatures of 100 K and 200 K, respectively. However, once we heated the structure to 270 K, only after 1 ps MD simulation, a complete transformation into the partially disordered mixed structure [5] of phase IV was observed. We then cooled down the system to 240 K, and after 3 ps MD simulations the structure transformed back into $C2/c$ structure. These MD simulations at 250 GPa are in excellent agreement with the experimental observation on the III → IV transition at ~280 K [2].

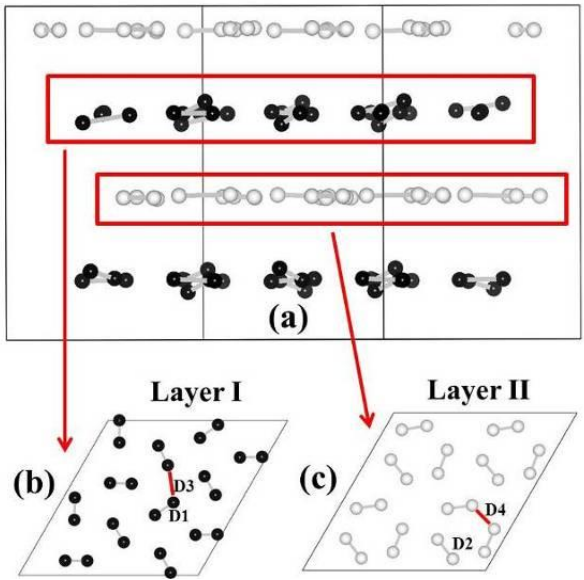

**Fig. 1 (color online)** (a) Side views of the mixed phase IV structure at 250 GPa. (b) Disordered $H_2$ layer (layer I) and (c) graphene-like layer containing three-molecule rings (layer II).

The mean square displacement (MSD) derived from our MD simulations as a function of time from 250 to 350 GPa at 300 K are shown in Fig. 2a. It is clearly seen that at 250 GPa, the MSD of hydrogen in layer I of phase IV remains a near constant with time, indicating the ultra-stability of molecules in layer I. In contrast, the MSD of hydrogen (black solid curve in Fig. 2a) in layer II grows evidently with time, signaling a novel feature of proton diffusion. From the slope of the MSD curve, an averaged diffusion coefficient of ~$0.3\times10^{-5}$ cm$^2$/s at 300 K was obtained at 250 GPa. With increasing pressure, it is found that the diffusion coefficient was dramatically increased and reached as high as ~$1.9\times10^{-5}$ cm$^2$/s at 350 GPa. Moreover, once we heated the system to 500 K at 250 GPa, more pronounced proton diffusion was evidenced with an unusually large diffusion coefficient at ~$2.1\times10^{-5}$ cm$^2$/s (blue curve in Fig. 2a). These proton diffusions at high pressures (e.g., 350 GPa) or at high temperatures (e.g., 500 K) are considerably large, even comparable to those for typical liquids, e.g., ~$2.3\times10^{-5}$ cm$^2$/s in liquid water at 298.16 K [15]. Our simulations thus provided the direct evidence on the quantum liquid-like state of phase IV and suggested a phase regime on the existence of liquid-like state over broad pressure and temperature ranges.

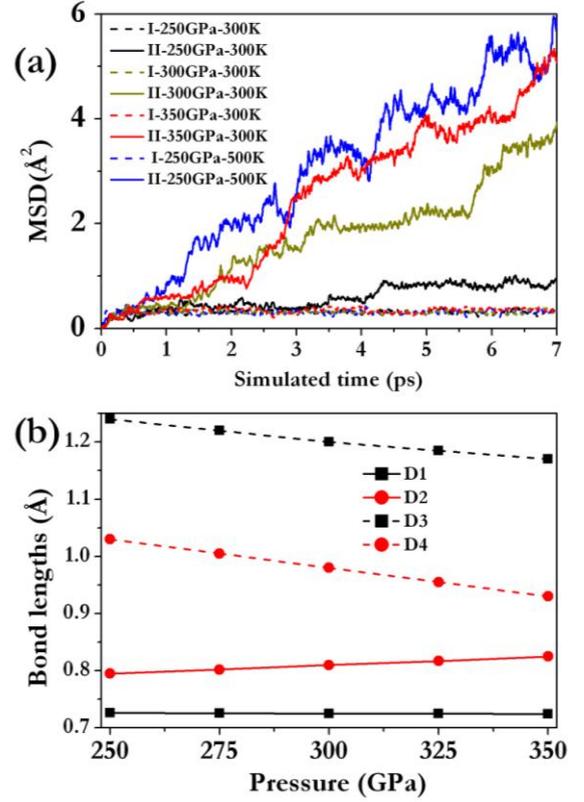

**Fig. 2 (color online)** (a) Mean squared displacements as a function of time at various pressures and temperatures for mixed phase IV are shown. (b) The averaged intra-molecular (D1 and D2 are shown in Fig. 1) and inter-molecular (D3 and D4 are shown in Fig. 1) bond lengths for mixed phase IV are plotted as a function of pressures. It is found that intra-molecular bond lengths (D2) in layer II gradually increase with pressure, while intra-molecular bond lengths (D1) in layer I remain nearly invariant with pressure.

We have also performed the variable-cell MD simulations for solid deuterium at 300 GPa. Similar diffusion effect in layer II is also found. However, the obtained diffusion coefficient (~$0.5\times10^{-5}$ cm$^2$/s) of deuterium is much less than that of hydrogen as expected due to the double heavier atomic mass of deuterium. Again, in agreement with hydrogen, we did not find any proton diffusion in layer I of solid deuterium. The trajectories of hydrogen protons in layers I and II are shown in Fig. 3. It is clearly seen that protons in layer II are readily to diffuse from one molecular site to its nearest-neighboring molecular site, but the diffusion is only confined within the *x-y* layers (no discontinuous coordinate changes were found in *z* direction).

To better illustrate the characteristic behaviors of hydrogen diffusion, two movies [16] were prepared by making use of hydrogen trajectories after 7.5 ps MD simulations at 300 K and 300 GPa. It is further confirmed that protons in molecules of layer II can readily transfer to the lattice sites of their neighboring molecules. More intriguingly, we found that the three-molecule rings in layer II can actually be rotating during the course of proton transfers. To rationalize such a rotation, we have calculated the rotational energy barriers at two pressures of 250 and 300 GPa, through particularly designed total-energy calculations. In the calculations, we gradually rotated the rings, but the center of rings remained unchanged. The resultant rotational energy barriers are small at 30 – 34 meV/atom (Fig. 4b). Such a low-energy barrier could lead to a rotation of rings. Under higher pressure, it was calculated that the rotational energy barrier becomes evidently smaller (Fig. 4b), leading to an easier rotation at elevated pressures. This is as expected since molecules in layer II under higher pressures are more elongated (Fig. 2b), resulting in a closer inter-molecular distance.

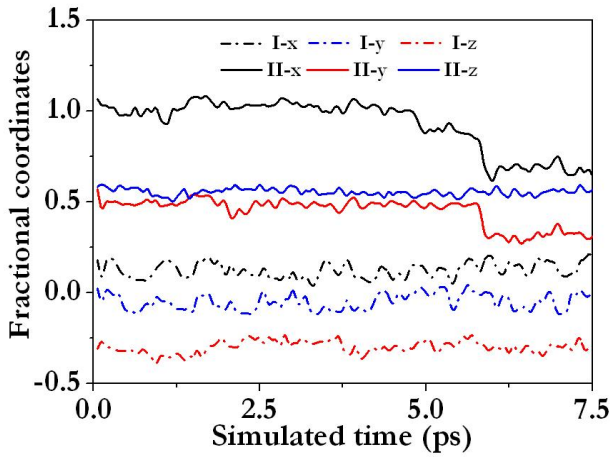

**Fig. 3 (color online)** The fractional coordinates of hydrogen protons in layers I and II for mixed phase IV at 300 GPa and 300 K as a functional of time, respectively.

In order to explore whether the unusual phenomena of proton diffusion of phase IV can occur in other molecular phases, we have performed additional variable-cell MD simulations for $Cmca$-4 structure at 350 GPa and 300 K and for $C2/c$ structure at 250 GPa and 200 K, respectively. Our simulations excluded the existence of proton transfers in these molecular phases (Fig. 4a). This is not unreasonable: (i) the intra-molecular bond lengths (~0.75 Å) in $C2/c$ structure at 250 GPa and those (~0.78 Å) in $Cmca$-4 structure at 350 GPa are evidently shorter than those (~0.8 Å) in layer II for phase IV at 250 GPa; (ii) the inter-molecular distances (~1.13 Å) in $C2/c$ structure at 250 GPa and those (~1.07 Å) in $Cmca$-4 structure at 350 GPa are longer than those (~1.03 Å) in layer II for phase IV at 250 GPa. By all appearance, we found the predicted proton transfer is intimately related to the particular elongation of hydrogen molecules in layer II of phase IV. The elongation leads to the stronger inter-molecular interaction and the weaker intra-molecular bondings.

Raman experiments [1, 2] observed that the Raman FWHM of $v_1$ vibron (corresponding to stretching mode of hydrogen molecules in layer II) in phase IV is considerably larger than that of phase III of solid hydrogen, and it rapidly increases with pressure. In contrast, the Raman FWHM of $v_2$ vibron (corresponding to stretching mode of hydrogen molecules in layer I) in phase IV has no obvious change with respect to that of phase III, and it remains nearly invariant with pressure. These experimental observations could be understood by our current simulations. Our simulated proton transfer in layer II of phase IV fundamentally revealed that molecules in layer II are short-lived, and thus hydrogen vibration is strongly anharmonic. These facts are the physical origin for the observed abnormally large Raman FWHM of $v_1$ vibron in phase IV. Furthermore, its large increase with pressure is attributed to the enhanced proton transfer at elevated pressures. In contrast, molecules in layer I remain ultra-stable, which are responsible for the little-affected Raman FWHM of $v_2$ vibron, even at higher pressures.

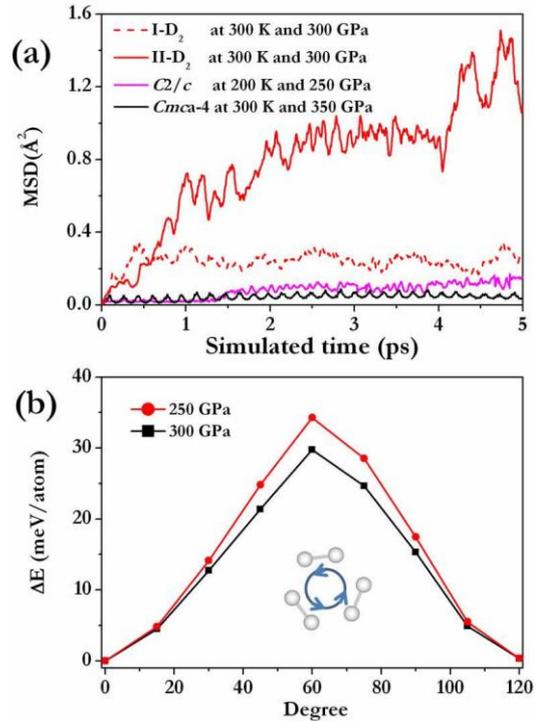

**Fig. 4 (color online)** (a) Mean squared displacements as a function of time at various pressures and temperatures for various phases ($C2/c$ and $Cmca$-4 phases for hydrogen, mixed phase IV for deuterium) are shown. (b) The total-energies of mixed phase I at 250 GPa and 300 GPa as a function of rotational angle, respectively. The lowest energies are scaled as the zero energy.

After the completion of our research, we were aware another work [17] that reported a similar *ab initio* variable-cell MD simulations on phase IV. Though they predicted the similar intra-layer hydrogen hoppings, our results are different to theirs in many aspects: (i) We did not find their predicted synchronized intra-layer atomic fluctuations changing layer I to layer II and *vice versa*, and thus the resultant indistinguishable distributions of hydrogen between layers I and II; (ii) Their predicted collaborative "hopping" of the hydrogen molecules in layer I was not found in our work. In our simulations, we revealed a neat picture on maintaining the mixed framework of layers I and II. Intra-layer proton transfer and rotation of rings (not found by them [17]) are limited to layer II. We do not understand the origin for these discrepancies, but a double larger simulation cell (48 moelcules) was used in our simulation. It is generally accepted that a larger supercell is necessary for MD simulations to efficiently reduce structural fluctuations [18, 19]. To eliminate the size influence on our simulation results, we have performed an additional variable-cell MD simulation at 300 GPa and 300 K with a much larger supercell containing 288 molecules by using a well-converged $2\times2\times2$ $k$-mesh. Again, we did not observe any structural fluctuations and the "hopping" of hydrogen molecules in layer I of phase IV after long (10 ps) variable-cell MD runs.

Our predicted proton transfer and the rotation of rings in layer II place a significant caution on any static total energy or phonon calculations based on harmonic approximation. Atoms in layer II of phase IV are intrinsically mobile, especially at higher pressures or higher temperatures before it melts. In this aspect, it is not unreasonable to regard phase IV as a proton conductor though conduction has been strictly confined within layer II. Specially designed high-pressure ion conductive measurements might be able to detect such an intriguing phenomenon.

In summary, our *ab initio* variable-cell MD simulations for phases III and IV of solid hydrogen and deuterium were able to identify the intra-layer proton transfer and the simultaneous rotation of three-molecule rings in layer II of phase IV. We found that proton transfer has a direct relation to the unique elongation of molecules in layer II. Our simulations are able to reveal that hydrogen molecules in layer II are short-lived and hydrogen vibration is strongly anharmonic. These facts are responsible for the experimental findings on the abrupt increase of $v_1$ Raman FWHM of phase IV. It is worth mentioning that the proton transfer becomes more pronounced with increasing pressure or temperature, and thus a phase regime on the existence of quantum liquid-like state of phase IV is highly expected in a wide range of high pressures and high temperatures.

Y.M. is highly grateful to the insightful discussions with Dr. Eugene Gregoryanz and Dr. Mikhail Eremets on phase IV. The authors acknowledge the funding supports from China 973 Program under Grant No. 2011CB808200, National Natural Science Foundation of China under Grant Nos. 11274136, 11025418 and 91022029, 2012 Changjiang Scholar of Ministry of Education, and Changjiang Scholar and Innovative Research Team in University (No. IRT1132). Part of calculations was performed in high performance computing center of Jilin University.

*Author to whom correspondence should be addressed: mym@jlu.edu.cn

References

[1] M. I. Eremets, and I. A. Troyan, Nature. Mater. **10**, 927 (2011).

[2] R. T. Howie, C. L. Guillaume, T. Scheler, A. F. Goncharov, and E. Gregoryanz, Phys. Rev. Lett. **108**, 125501 (2012).

[3] C. J. Pickard, and R. J. Needs, Nature. Phys. **3**, 473 (2007).

[4] C. J. Pickard, M. Martinez-Canales, and R. J. Needs, Phys. Rev. B **85**, 214114 (2012).

[5] H. Liu, L. Zhu, W. Cui, and Y. Ma, J. Chem. Phys **137**, 074501 (2012).

[6] S. Biermann, D. Hohl, and D. Marx, Solid State Commun. **108**, 337 (1998).

[7] E. R. Hernandez, J.Chem. Phys. **115**, 10282 (2001).

[8] G. Kresse, and J. Hafner, Phys. Rev. B **47**, 558 (1993).

[9] E. R. Hernandez, A. Rodriguez-Prieto, A. Bergara, and D. Alfe, Phys. Rev. Lett. **104**, 185701 (2010).

[10] G. Kresse, and J. Furthmüller, Phys. Rev. B **54**, 11169 (1996).

[11] G. Kresse, and D. Joubert, Phys. Rev. B **59**, 1758 (1999).

[12] P. E. Blochl, Physical Review B **50**, 17953 (1994).

[13] J. P. Perdew, K. Burke, and M. Ernzerhof, Phys. Rev. Lett. **77**, 3865 (1996).

[14] H. J. Monkhorst, and J. D. Pack, Phys. Rev. B **13**, 5188 (1976).

[15] K. Krynicki, C. D. Green, and D. W. Sawyer, Faraday Discussions of the Chemical Society **66**, 199 (1978).

[16] Reserved for EPAPS.

[17] A. F. Goncharov *et al.*, arXiv:1209.3895 (2012).

[18] M. A. Morales, C. Pierleoni, E. Schwegler, and D. M. Ceperley, Proc. Natl. Acad. Sci. U.S.A. **107**, 12799 (2010).

[19] W. Lorenzen, B. Holst, and R. Redmer, Physical Review B **82**, 195107 (2010).